\documentclass{moriond}

\def\Journal#1#2#3#4{{#1} {\bf #2}, #3 (#4)}

\def\PRL{\em Phys. Rev. Lett.}
\def\PRD{{\em Phys. Rev.} D}

\def\be{\begin{equation}}
\def\ee{\end{equation}}
\def\bea{\begin{eqnarray}}
\def\eea{\end{eqnarray}}

\newcommand{\Photo}{}

\begin{document}
\vspace*{4cm}
\title{THE MUON G-2 EXPERIMENT AT FERMILAB}

\author{ K.R. LABE }

\address{on behalf of the Muon $g-2$ Collaboration\\ Department of Physics, Cornell University, \\
Ithaca, New York, 14853, United States}

\maketitle\abstracts{The muon magnetic anomaly, $a_{\mu}$, is a powerful test of the Standard Model
of particle physics.  A new experiment at Fermilab has recently measured $a_{\mu}$ with
unprecedented precision, confirming the results of the earlier Brookhaven experiment and
strengthening the tension with the prediction of the Standard Model as determined by dispersive
methods.  We here describe the experimental technique, recapitulate the recent result, and 
discuss some of the improvements made for subsequent analyses.}

\section{Background}
The magnetic moment $\vec{\mu}$ of a fundamental particle with charge $q$, mass $m$, and
spin $\vec{S}$ is given by
\begin{equation}
\vec{\mu} = g \frac{q}{2m}\vec{S} \, .
\end{equation}
The dimensionless parameter $g$, the gyromagnetic ratio, describes the overall strength of the
magnetic moment in units of the classical magnetic moment.

Since 1948, it has been
understood \cite{schwinger} that the gyromagnetic ratio of electrons
(and also muons) differs 
from the Dirac Model expectation of 2 at the per-mille level due to interactions of the leptons 
with virtual particles.
This additional contribution is called the magnetic anomaly, $a$, defined $a \equiv \frac{g-2}{2}$.
The contributions of the known
Standard Model particles to the magnetic anomaly can be calculated very precisely,
making this quantity an outstanding test of the Standard Model.

An earlier measurement \cite{bnl} of the muon magnetic moment, carried out at Brookhaven National
Laboratory, was discrepant from the theory prediction by roughly 3$\sigma$.  This
motivated the construction of a new experiment at Fermi National Accelerator Laboratory to repeat
the measurement at higher precision.  The first result \cite{r1} of the new experiment, which agrees
with the measurement at Brookhaven, is discussed here together with the improvements expected
for future publications.
The high precision expected from the new experiment also motivated an extensive effort in the theory
community to further improve the understanding of the prediction within the Standard Model.
Recently, the Muon $g-2$ Theory Initiative \mbox{published \cite{whitepaper}} a determination of this
quantity with an
uncertainty of 0.37 ppm.  An alternative determination \cite{bmw} based on lattice QCD stands in
some tension with the dispersive estimate of the Theory Initiative.

\section{Experimental Principle}
In the presence of an externally applied magnetic field $\vec{B}$, a charged particle will traverse a 
circular orbit, and its spin will feel a torque
\begin{equation}
\vec{\tau} = \vec{\mu} \times \vec{B} \, .
\end{equation}
Assuming the spin axis is not aligned with the magnetic field, this torque will cause the spin axis
to precess about the magnetic field direction.  By measuring the frequency of this spin precession,
$\omega_s$, 
and the strength of the applied magnetic field, the magnetic moment may be deduced.

In our experiment, the spin precession is determined through the influence of the spin orientation on
the kinematics of the muon decay.  We observe the Michel decays of an ensemble of spin-polarized
muons into positrons (and neutrinos which escape detection).  The differential cross section
for this decay strongly 
correlates the momentum of the daughter positrons with the muon spin direction in the muon rest
frame.  As the muon spin direction precesses in the magnetic field with respect to its momentum
(which is also rotating with cyclotron frequency $\omega_c$), the
mean boost given to the daughter positrons, and thus their mean energy, also varies harmonically.
The anomalous precession frequency ($\omega_a \equiv \omega_c - \omega_s$) can therefore be
measured through the rate of variation of the positron energy in the laboratory frame.

The magnetic field strength is measured using a similar spin precession technique, but in this case,
the nuclear spin precession of petroleum jelly samples is measured using nuclear magnetic resonance
(NMR) techniques.  This allows the well-known nuclear magneton to serve as a comagnetometer
for the muons experiencing the same field.

The experimental apparatus consists of a 3.56 m radius, 1.45 T superferric magnetic storage ring
used to confine the
3.1 GeV muon beam provided by the Fermilab accelerator in 16 bunches that arrive every 1.4 s.
Vertical confinement is provided by a set of four electrostatic quadrupoles.  The apparatus is
instrumented with detectors of three types.  First are 24 electromagnetic calorimeters spaced around
the interior of the magnetic ring used to measure the positron energies.  Each calorimeter consists of
a segmented 9 $\times$ 6 array of PbF$_2$ crystals, each provided with a SiPM digitized at 800 MSPS.  Second are two straw tracking stations located within the storage ring vacuum but outside
the muon beam path.  The straw trackers measure positron momenta from which the muon beam
dynamics can be determined.  Third are the nuclear magnetic resonance probes used to measure
the magnetic field strength.

\section{Overview of the Run-1 analysis}

The first experimental run of the experiment was conducted between March and June 2018, and
results were published in April 2021 \cite{r1}.  The analysis consists of three parts, discussed in turn
below: the measurement
of the muon anomalous precession frequency $\omega_a$ \cite{wa}, corrections to the measured
anomalous precession frequency due to
muon beam dynamics effects \cite{beam}, and the measurement of the magnetic field strength
$\omega_p$ \cite{wp}.  Combining these results with external reference measurements $k$, 
$a_{\mu}$ was determined by
\begin{equation}
a_{\mu} = k\frac{\omega_a}{\omega_p}g_e \, .
\end{equation}

A blind analysis was conducted by multiple independent analysis groups on each of four subsets of 
the Run-1 data.  After establishing that the independent analyses were consistent, the results were
combined and unblinded.  We found $a_{\mu} = 116 592 040 \pm 54 \times 10^{-11}$.  This agreed
with the measurement made at Brookhaven at a level of 0.6$\sigma$.  When combined with the
Brookhaven measurement, the world average disagrees with the Standard Model prediction
\cite{whitepaper} using dispersive techniques at 4.2$\sigma$.  The agreement with the
lattice QCD estimate \cite{bmw} is significantly better.  A summary of the experimental error budget
is presented in Table \ref{tab:errors}.

\begin{table}
\caption{Error budget of the Run-1 analysis of the muon magnetic anomaly.}
\begin{center}
\begin{tabular}{|cc|}
\hline
Quantity & Uncertainty (ppb) \\ \hline
Precession (stat) & 434 \\
Precession (syst) & 58 \\
E Field Correction & 53 \\
Phase-Acceptance & 75 \\
Magnetic Field & 56 \\
Kicker Field Transient & 37 \\
Quad Field Transient & 92 \\
External Factors & 25 \\
Total & 462 \\
\hline
\end{tabular}
\end{center}
\label{tab:errors}
\end{table}

\subsection{Anomalous precession analysis and corrections}
To determine the anomalous precession frequency, the number of high-energy positrons entering the 
calorimeters is counted as a function of time in the storage ring.  A time series of these high-energy
positrons has a characteristic oscillation with the anomalous precession frequency.  In practice,
a slightly
more complex analysis with higher precision is used to obtain the time series by weighting
the positrons according to the decay asymmetry associated with their energy.

Besides the anomalous precession oscillation, there are a number
of other features present in the time series that must be correctly fit in order to obtain an
unbiased frequency estimate.  These include the exponential decay of the muon population,
acceptance effects due to the coherent betatron oscillations of the muon beam in the storage ring, and
the mechanical losses of muons from the storage ring before decay.

Once these effects are included in the analysis, it is possible to achieve an excellent
$\chi^2$ to the fitting function describing the data over many muon lifetimes.  Many additional checks
on the self-consistency of the analysis were performed, including confirming that the fit residuals were
without time structure and that the fit results were consistent across calorimeters.  Furthermore, six
independent analyses using different reconstructions and analysis techniques all yielded consistent
results.

The anomalous precession frequency obtained from the time series analysis must be corrected for
a small
number of effects that bias the measurement.  These include an electric field correction needed due to
the electrostatic quadrupoles (which appear as a motional magnetic field to the muons), a pitch
correction to account for the muons' motion out of the plane due to their vertical oscillations,
and phase-acceptance and muon loss corrections that reflect time dependence in the mean
accepted spin phase.

\subsection{Magnetic field analysis}
The magnetic field in the muon storage volume can be measured with exquisite precision with 17
NMR probes mounted in a moveable trolley carriage that can be pulled through the storage ring.
These trolley probes can measure many field multipole moments with high azimuthal resolution, and
these probes are calibrated against an absolute reference probe.  However, the trolley cannot
be left in the storage ring during data taking; it is therefore used only two to three times per week
during dedicated trolley runs.

In the intervening time, the evolution of the magnetic field multipole moments is tracked with a set
of 378 NMR probes located at 72 azimuthal locations just above and below the storage volume.
These probes are calibrated against the trolley during the trolley runs and interpolate the
field moments between the trolley measurements.

From this set of time-dependent field moments, the average magnetic field strength experienced by
the muons in the anomalous precession analysis can be determined by weighting the field at each
time slice
according to the number of muons then observed, and by convolving the multipole moments
of the field
with the multipole moments of the muon beam distribution, as measured with the straw tracker
system.

\section{Future improvements}
\label{sec:improve}
The collaboration is currently working on analyzing the data collected in the second and third 
experimental runs.  Improvements to the most significant uncertainties reported in
Table \ref{tab:errors} are anticipated.  In this section, we discuss the reasons for these improvements.

\subsection{Statistics}
In the first experimental run, only 6\% of the final statistical goal was met.  The two subsequent
experimental runs were longer
and together represent about four times the number of muons collected in the first run.  We therefore
expect the statistical uncertainty, which dominated the Run-1 result, to be reduced by roughly a factor
of two.

The experiment is currently collecting its fifth experimental run and is on track to meet its final goal
of 100 ppb statistical uncertainty.

\subsection{Replaced Quadrupole HV resistors}
A significant difficulty in analyzing the Run-1 data was caused by damaged high-voltage resistors used
in charging the quadrupole plates.  These quadrupole plates are pulsed in time with the muon
bunches, and each of the 32 plates is regulated by its own resistor.  During the first experimental run,
two of these resistors were damaged in such a way as to significantly increase the charging time of
the connected plates.  This caused an asymmetric time dependence to the vertical focusing that subsequently affected a number of beam characteristics such as its position and width.
It also dominated the
phase-acceptance effect which contributed a significant systematic uncertainty to the analysis.
Following Run-1, the problematic resistors were replaced, which will
significantly reduce the uncertainty associated with the phase-acceptance effect.

\subsection{Quadrupole Field Transient}
Two field transients impact the muon spin evolution but cannot be tracked using the NMR probe 
system because of the high frequency of the transients and the shielding of the NMR
probes.  These transients are instead measured with dedicated probes and appropriate corrections
applied to the field measurement.

One such transient comes from the pulsing of the quadrupole plates, which induces mechanical
vibrations.  For the Run-1 analysis, this transient was measured with special
NMR probes inserted into the quadrupoles at a limited number of positions.  The low granularity of
the measurement and uncertainty about the stability of the effect over time limited the precision of the
correction.  Since that time, an improved set of NMR probes on a trolley frame that can be pulled 
through the storage ring has been deployed, allowing a highly detailed measurement to be carried out.
These data will allow for a significant reduction in the uncertainty on the correction in the future.

\subsection{Stronger Kick}
A non-ferric kicker magnet is used to place the injected muon bunch onto a stable orbit in the storage
ring \cite{kicker}.  For Run-1, this kicker system was unable to operate at a high enough voltage to
optimize the storage of muons with the so-called ``magic" momentum, at which the electric field
correction is minimized.  Further upgrades to the kicker circuit and supporting systems enabled
the kicker to reach its design targets in Run-3.  This improves the number
of muons stored, reduces the size of the electric field correction, and reduces the amplitude of the
coherent betatron oscillations, which impact the anomalous precession analysis.

\section{Conclusion}
The first result of the Muon $g-2$ experiment at Fermilab \cite{r1} has confirmed the tension
with the Standard Model prediction calculated using dispersive techniques \cite{whitepaper} first
established at the Brookhaven experiment \cite{bnl}.  Currently the analysis of the data from the
second and third year of the experiment is underway, with significant improvements expected, as
described in Section \ref{sec:improve}.  We anticipate the ongoing fifth experimental run will
conclude the $\mu^+$ program, with a new measurement of $a_{\mu}$ for the negative muon
to commence later this calendar year.

\section*{Acknowledgments}

This work was supported in part by Fermilab and the US DOE under contract \mbox{No.\ DE-SC0008037}.


\section*{References}

\begin{thebibliography}{99}
\bibitem{schwinger} J.S. Schwinger, \Journal{{\em Phys. Rev.}}{73}{416}{1948}.
\bibitem{bnl} G.W. Bennett {\it et al}, \Journal{\PRD}{73}{072003}{2006}.
\bibitem{r1} B. Abi {\it et al}, \Journal{\PRL}{126}{141801}{2021}.
\bibitem{whitepaper} T. Aoyama {\it et al}, \Journal{{\em Phys. Rep.}}{887}{1}{2020}.
\bibitem{bmw} S. Borsanyi {\it et al}, \Journal{{\em Nature}}{593}{51}{2021}.
\bibitem{wa} T. Albahri {\it et al}, \Journal{\PRD}{103}{072002}{2021}.
\bibitem{beam} T. Albahri {\it et al}, \Journal{{\em Phys. Rev. Accel. Beams}}{24}{044002}{2021}.
\bibitem{wp} T. Albahri {\it et al}, \Journal{{\em Phys. Rev.} A}{103}{042208}{2021}.
\bibitem{kicker}  A.P. Schreckenberger {\it et al}, \Journal{{\em Nucl. Instrum. Meth.} A}{1011}{165597}{2021}.

\end{thebibliography}
\end{document}